# Applying Social Network Analysis to Analyze a Web-Based Community

Mohammed Al-Taie
Master of computer Science and Communication dept.
Arts, Sciences and Technologies University
Beirut, Lebanon

Seifedine Kadry
Master of computer Science and Communication dept.
Arts, Sciences and Technologies University
Beirut, Lebanon
Email: skadry@gmail.com

*Abstract*— this paper deals with a very renowned website (that is Book-Crossing) from two angles: The first angle focuses on the direct relations between users and books. Many things can be inferred from this part of analysis such as who is more interested in book reading than others and why? Which books are most popular and which users are most active and why? The task requires the use of certain social network analysis measures (e.g. degree centrality).

What does it mean when two users like the same book? Is it the same when other two users have one thousand books in common? Who is more likely to be a friend of whom and why? Are there specific people in the community who are more qualified to establish large circles of social relations? These questions (and of course others) were answered through the other part of the analysis, which will take us to probe the potential social relations between users in this community. Although these relationships do not exist explicitly, they can be inferred with the help of affiliation network analysis and techniques such as m-slice.

Book-Crossing dataset, which covered four weeks of users' activities during 2004, has always been the focus of investigation for researchers interested in discovering patterns of users' preferences in order to offer the most possible accurate recommendations. However; the implicit social relationships among users that emerge (when putting users in groups based on similarity in book preferences) did not gain the same amount of attention. This could be due to the importance recommender systems attain these days (as compared to other research fields) as a result to the rapid spread of e-commerce websites that seek to market their products online.

Certain social network analysis software, namely Pajek, was used to explore different structural aspects of this community such as brokerage roles, triadic constraints and levels of cohesion.

Some overall statistics were also obtained such as network density, average geodesic distance and average degree.

1. INTRODUCTION

Social network analysis (SNA) is concerned with realizing the linkages among social entities and the implications of these linkages [33].

It has evolved due to the synergy of three fused (separated, in sometimes) strands. These three strands were formed from the efforts of sociometric analysts who worked on small groups and came up with technical advances in methods of graph theory, the Harvard researchers of the 1930s who discovered patterns of interpersonal relations and the formation of cliques, and the Manchester anthropologists who investigated the structure of community relations in tribal and village societies [28].

The essential goal of SNA is to examine relationships among individuals, such as influence, communication, advice, friendship, trust etc., as researchers are interested in the evolution of these relationships and the overall structure, in addition to their influence on both individual behavior and group performance [29].

As for [23], they conducted a research to measure the growth of SNA field for the period (1963-2000). They consulted three databases that related to three branches of science (namely sociology, medicine and psychology). Among their findings were that the real growth of the field began in 1981 and there was no sign of decline and that the development in the field began in sociology faster than what it was in medicine and psychology. They noticed that the success which SNA has witnessed in the eighties was due to the institutionalization of social network analysis since late seventies and the recent availability of textbooks and software packages.

Today, social network analysts have an international organization called '*The International Network for Social Network Analysis*' or INSNA, which holds annual meetings and issues a number of professional journals. Also, a number of centers for network searching and training have opened worldwide [8].

2. APPLICATIONS OF SOCIAL NETWORK ANALYSIS

SNA is involved in a many tasks, such as identifying most important actors in a social network through the use of centrality analysis, community detection, identifying the role associated with each member through conducting role analysis, network modeling for large-scale complex networks, how the information diffuses in a network and viral marketing [31].

## 2.1 Semantic Web

The idea of semantic web is to implement advanced knowledge techniques to fill the gap between machine and human. This implies providing the required knowledge that enables a computer to easily process and reason [21].

As for [7], he merged the semantic web frameworks model (which allows representing and exchanging knowledge across web application) and SNA model (which proposes graph algorithms to characterize the structure of a social network and its strategic positions). This combination was necessary in order to go beyond mining the flat link structure of social graphs.

## 2.2 Social recommendation systems

The use of SNA in the field of designing recommender systems (RS) is still in primitive stages [36]. However; it is expected that new methods using SNA will be incorporated in recommender system design [24], [36].

For [15], they presented a collaborative-based recommendation system that uses trading relationships to calculate level of recommendation for trusted online auction sellers. They used k-core, center weights algorithms and two social network indicators to create a recommender system that could suggest risks of collusion associated with an account.

## 2.3 Software development

Social network analysis in software engineering plays an important role in project support as more projects have to be conducted in globally-distributed settings.

In [16], they developed a method and a tool implementation to apply SNA techniques in distributed collaborative software development, as this provides surpassing information on expertise location, coworker activities and personnel development.

As for [20], they applied SNA to code churn information, as an additional means to predict software failures. Code churn is a software development artifact (common to most large projects) and can be used to predict failures at the file level. Their goal was to examine human factor in failure predicting. They conducted their case study on a large Nortel networking product, comprising more than 11000 files and three million lines of code.

## 2.4 Health

Network analysis, more and more, is becoming well-known in infectious disease epidemiology, such as Human Immunodeficiency Virus (HIV) and Sexually Transmitted Diseases (STD). Also, a strong trend is emerging towards using inter-organizational network analysis to detect patterns of health care delivery such as service integration and collaboration [13].

For [26], they conducted a study to know the relationship between SNA and the epidemiology and prevention of STD. They argue that SNA will be of a great utility in the study of STD.

As for [10], they found that the traditional contact tracing (the technique which they used at the beginning of their search to discover the reason behind the spread of tuberculosis in a medium-size community in British Colombia) did not identify the source of the disease. By using whole-Genome sequencing and SNA, they discovered that the cause was related to socio-environmental factor.

## 2.5 Cybercrimes

Cybercrimes are offences that are committed against individuals or groups of individuals with a criminal motive using modern telecommunication networks such as Internet and mobile phones [35].

Ref. [37] presented a framework to analyze and visualize weblog social networks. A weblog is a website where the contents are formulated in a diary style and maintained by the blogger. This environment makes a good platform for organizing crimes. With the ability to analyze and visualize weblog social networks in crime-related matters, intelligence agencies will have additional techniques to secure the society.

To investigate hacker community, [18] examined the social structure of an unknown hacker community called 'Shadowcrew'. For the investigation, they used text mining and network analysis to discover the relationships among hackers. Their work showed the decentralized composition of that community. Based on that analysis, they found that this community exhibits features of deviant team organization structure.

## 2.6 Business

SNA applies to a wide range of business fields, including human resources, knowledge management and collaboration, team building, sales and marketing and strategy.

Ref. [12] looked at SNA as a tool which can enhance the empirical quality of Human Resource Development (HRD) theory in areas such as organizational development, organizational learning, etc. He argues that SNA will add much to HRD fields by measuring the relations between individuals, and the effect those relations have on human capital output.

For [6], they studied the influence of SNA and sentiment analysis in predicting business trends. They focused on predicting the successes of new movies, in the box office, for the first four weeks. They were trying to predict prices on the Hollywood Stock Exchange (HSE), and the ratio of gross income to the budget of the production. They depended on data posts from Internet Movie Database (IMDb) forums to get sentiment metrics for positivity and negativity based on forum discussions.

Through using a Twitter dataset, [38] tried to predict stock market indicators such as Dow Jones, S&P500 and NASDAQ. They took about one hundredth of the total Twitter data that covered six months of activity. Through analyzing the relationship between data and stock market indicators, they found that emotional tweets displayed negative correlation to NASDAQ and S&P500, but gave positive correlation to VIX. They concluded that Twitter analysis can be used as a tool to predict stock market of the next day.

## 2.7 Collaborative Learning

Social network analysis provides meaningful and quantitative insights into the quality of knowledge construction process. It can effectively assess the performance of knowledge building process.

Ref. [27] showed that concepts of SNA, adapted to the collaborative distant-learning, can assist measuring small group cohesion. Their data were taken from distance-learning experiment of ten weeks. They used different ways to measure cohesion in order to highlight active subgroups, isolated people and roles of the members in the group communication structure. They argue that their method can show global attributes at the group level and individual level, and will help the tutor in following the collaboration in the group.

Ref. [25] has investigated the potential use of SNA to evaluate programs that seek to enhance school performance through encouraging greater collaboration among teachers. Through gathering data about teacher collaboration in schools, they mapped the distribution of expertise and resources needed to achieve reforms. One of their findings was that although the majority of teachers consider collecting social network data to be feasible, other teachers show concerns related to privacy and data sharing.

### 3. GRAPH THEORY

The origins of graph theory can be traced back to Euler's work on the Konigsberg bridges problem (1735), which subsequently led to the concept of an eulerian graph. The study of cycles on polyhedra by the Revd. Thomas Penyngton Kirkman (1800-95) and Sir William Rowan Hamilton (1805-05) led to the concept of a Hamiltonian graph [11].

The simplest definition of a graph is that it is a set of points and lines connecting some pairs of the points. Points are called 'vertices', and lines are called 'edges'. A graph $G$ is a set $X$ of vertices together with a set $E$ of edges and it is written as: $G = (X, E)$.

For a given vertex ($x$), the number of all vertices adjacent to it is called 'degree' of the vertex $x$, denoted by $d(x)$. The maximum degree over all vertices is called the maximum degree of $G$, denoted by $\Delta(G)$.

The adjacent vertices are sometimes called neighbors of each other, and all the neighbors of a given vertex $x$ are called the neighborhood of $x$. The neighborhood of $x$ is denoted by $N(x)$. The set of edges incident to a vertex $x$ is denoted by $E(x)$.

One can describe a graph by giving just the list of all of its edges. For graph $G$, the edge list, denoted by $J(G)$ is the following:

$$J(G) = \{\{x_1,x_2\},\{x_2,x_3\},\{x_3,x_4\},\{x_4,x_5\},\{x_1,x_5\},\{x_2,x_5\},\{x_2,x_4\}\}.$$

A loop is an edge connecting a vertex to it-self. If a vertex has no neighbors, i.e. its degree is 0, then these vertices are said to be isolated. If there are many edges connecting the same pair of vertices, then these edges are called 'parallel' or 'multiple'. A simple adjacency between vertices occurs when there is exactly one edge between them.

In a graph, an ordered pair of vertices is called an 'arc'. If $(x,y)$ is an arc, then $x$ is called the initial vertex and $y$ is called the terminal vertex. A graph in which all edges are ordered pairs is called the 'directed graph', or 'digraph'.

Graphs in which order is not important are called 'undirected graphs'. Undirected graphs without loops and multiple edges are called 'simple graphs' or just simply 'graphs'.

A graph in which all vertices can be numbered $x_1, x_2, \ldots, x_n$ in such a way that there is precisely one edge connecting every two consecutive vertices and there are no other edges, is called a 'path', while the number of edges in a path is the 'length'.

A graph is called 'connected' if in it any two vertices are connected by some path; otherwise it is called 'disconnected'. It means that in a disconnected graph there always exists a pair of vertices having no path connecting them. Any disconnected graph is a union of two or more connected graphs; each such connected graph is then called a 'connected component' of the original graph. A 'cycle' is a connected graph in which every vertex has degree 2. It is denoted by $C_n$ where $n$ is the number of vertices.

A simple adjacency between vertices occurs when there is exactly one edge between them. A graph in which every pair of vertices is an edge, is called 'complete', denoted by $K_n$ whereas usually, $n$ is the number of vertices. It is complete because we can't add any new edge to it and obtain a simple graph.

If we have a graph $G = (X,E)$ and a vertex $x \epsilon X$. The deletion of $x$ from $G$ means removing $x$ from set $X$ and removing from $E$ all edges of $G$ that contain $x$. However, the deletion of an edge is easier than that of the vertex, as it comprises only removing the edge from the list of edges.

Let $G = (X,E)$ be a graph, $x,y \epsilon X$. The distance from $x$ to $y$, denoted by $d(x,y)$, is the length of the shortest $(x,y)$-path. If there is no such path in $G$, then $d(x,y) = \infty$. In this case, $G$ is disconnected and $x$ and $y$ are in different components.

The diameter of $G$ denoted by $diam(G)$ is $\max_{x,y \epsilon X} d(x,y)$, which means it is the distance between the farthest vertices.

A graph $G = (X,D)$ is called 'weighted' if each edge $D \epsilon D$ is assigned a positive real number $w(D)$ called the weight of edge $(D)$. In many practical applications, the weight represents a distance, cost, time, capacity, probability, resistance, etc.

In a graph $G$, a walk is an alternating sequence of vertices and edges where every edge connects preceding and succeeding vertices in the sequence. It starts at a vertex, ends at a vertex and has the following form: $x_0, e_1, x_1, e_2, \ldots, e_k, x_k$.

A digraph $N = (X,A)$ is called a 'network', if $X$ is a set of vertices (also called nodes), $A$ is a set of arcs, and to each arc $a \epsilon A$ a non-negative real number $c(a)$ is assigned which is called the capacity of arc $a$. For any vertex $y \epsilon X$, any arc of type $(x,y)$ is called 'incoming, and every arc of type $(y, z)$ is called outcoming.

A digraph is (weakly) connected if its underlying graph is connected. A digraph is strongly connected if from each vertex to each other vertex there is a directed walk.

A cut-vertex (or cutpoint) is a vertex whose removal increases the number of components. A cut-edge is an edge whose removal increases the number of components [32].

## 4. CASE STUDY: BX-DATASET USING SNA

### 4.1 Data Description

Our dataset, which is available for free download from the internet, has two types of file extension: the (.sql) format and the (.csv) format. Three files are extracted when dealing with the second type of data files: BX-Books, BX-Users and BX-Book-Ratings. The BX-Books file contains information about the books available in the website database. The BX-Users file contains demographic information about registered users, namely location and age. The BX-Book-Ratings file contains the relational data that connect between users and rated items, in addition to the weight of the relationship (expressed as a numerical value on a scale from 0 to 10).

The BX dataset was collected in a 4-week crawl (August/September 2004) by [40] from the Book Crossing, a community where users around the world exchange information about books.

The dataset contains 1,149,780 implicit and explicit ratings on a scale from 0 to 10. Implicit ratings are expressed by 0 on the scale and constitute 716,109 ratings. The remaining 433,681 ratings are regarded as explicit ratings across 1 to 10 on the scale. The total number of users is 278,858 and of the books is 271,379 [30].

Ref. [14] suggest that BX dataset also contains many more implicit preferences, like when users buy books but they do not explicitly rate them, which gives a positive indication towards those books.

BX dataset suffers, like any other public dataset, from a number of drawbacks such as low density of user ratings; a problem makes predictions so noisy in that context. This issue was treated by other researchers through taking only a subset of the BX-dataset [4]. The demographic information contains what it looks erroneous and incomplete data. Also, if the dataset were to have more demographic information (such as gender or occupation) we would have had more deep understanding of users' preference.

Ref. [39] has discretized the BX-dataset into five general domains (based on content):

TABLE I. BOOK DOMAINS IN BX-DATASET

| Domain #1 | Domain #2 | Domain #3 | Domain #4 | Domain #5 |
|---|---|---|---|---|
| Mystery and Thrillers | Science Fiction and Fantasy | Science | Business and Investing | Religion and Spirituality |

### 4.2 Data Pre-processing

Removing implicit ratings (those with value=0 on the scale) was necessary since implicit ratings are written reviews rather than numerical values. So, from the original dataset which comprised 1,149,780 ratings, we are left now only with 433,659 ratings (i.e. on a scale from 1 to 10).

### 4.3 Software

The specific software which we used in our analysis was Pajek, a program for analysis and visualization of large networks [1]. Several reasons stood behind the use of this software: Pajek is capable of dealing with large networks (several hundred thousand and even millions of nodes), a task not every program can handle successfully. It is freely available to download from the internet. It has a simple GUI, which gives the space for machine resources to function easily and efficiently. It has a well-illustrated user's manual and a lot of free compatible datasets for testing purposes. It has powerful visualization tools and several data analytic algorithms. It has the ability to deal with different types of networks and many networks at the same time. Also, Pajek has the ability to engage with very powerful statistical analysis tools (R and SPSS). The software release we used was 2.05.

### 4.4 Two-Mode Network Analysis

A two-mode network data contain measurements on which actors from one of the sets have ties to actors in the other set. Actors in one of the sets are senders, while those in the other are receivers [33]. Examples of two-mode networks include corporate board management, attendance at events, membership in clubs, participation in online groups, membership in production teams and even course-taking patterns of high school students [2].

#### 4.4.1 Mother Network Analysis

The first network that we analyzed was the mother network (a name we used to describe the network that covers the entire scale of ratings, i.e. from 1 to 10).

Analyzing this network helped us answering the question: which users have made the highest number of ratings (most active users)? We were also able to answer the question: which books obtained the highest number of ratings (no matter whether they were negative or positive)? Let's take a look at some of the overall statistics, evaluated using Pajek:

TABLE II. OVERALL STATISTICS OF THE MOTHER NETWORK

| Metric | Value |
|---|---|
| Graph Type | Directed |
| Dimension | 263631 |
| Number of Arcs | 433660 |
| Network Density | 0.00000624 |
| Number of Loops | 0 |
| Number of Multiple Lines | 0 |
| Average Degree | 3.28990142 |
| Connected Components | 14684 |
| Single-Vertex Connected Component | 0 |
| Maximum Vertices in a Connected Component | 229036 (86.877%) |

It is a directed two-mode network with density equals 0.00000624, which is very low. Network dimension is 263631 and the number of ties is 433660 (the more number of nodes in a network, the less network density). The network has neither loops nor multiple lines and the average degree is 3.28990142. The number of connected components is 14684, which is very high (due to the high dispersion in users' choices) and the largest component consists of 229036 nodes.

The network has no isolated vertices. The importance of identifying the largest component (also called giant component) in a community is that it helps measuring the effectiveness of the network at doing its job [22].

The highest and lowest out-degrees and out-degree centralization values of the mother-network were as follows:

TABLE III. HIGHEST AND LOWEST OUT-DEGREES AND OUT-DEGREE CENTRALIZATION OF THE MOTHER NETWORK

| Metric | Value | Frequency |
|---|---|---|
| Highest output degree value | 8522 | 1 |
| Lowest output degree value | 1 | 45375 |
| Network out-degree Centralization | 0.03231949 | - |

We can see that only one node obtained the highest number (8522) of outgoing ties (most active user) from among 263631 nodes, and that 45375 other nodes (approximately 1/6 of network nodes) supplied only 1 vote (least active users). The analysis also gave us 185833 nodes with zero out-degree (not shown in the table above). This is because Pajek analyzed both types of nodes, namely users and books, and that the nodes with out-degree=0 represent books (destination of relation). The highest ten out-degree values (representing most active users) of the mother- network were as follows:

TABLE IV. HIGHEST TEN OUT-DEGREE VALUES (MOST ACTIVE USERS) IN THE BX-DATASET

| Rank | Out-Degree | Normalized Out-Degree | User ID | Age | Country |
|---|---|---|---|---|---|
| 1. | 8522 | 0.0323 | 11676 | Null | N/A |
| 2. | 5802 | 0.0220 | 98391 | 52 | USA |
| 3. | 1969 | 0.0075 | 153662 | 44 | USA |
| 4. | 1906 | 0.0072 | 189835 | Null | USA |
| 5. | 1395 | 0.0053 | 23902 | Null | UK |
| 6. | 1036 | 0.0039 | 76499 | Null | USA |
| 7. | 1035 | 0.0039 | 171118 | 47 | Canada |
| 8. | 1023 | 0.0039 | 235105 | 46 | USA |
| 9. | 968 | 0.0037 | 16795 | 47 | USA |
| 10. | 948 | 0.0036 | 248718 | 43 | USA |

Some users have higher out-degree values than others since they have provided a higher number of book ratings; in other word they are more active than their associates. We can see that 70-80% of the people whose outgoing links were probed were from USA, and that the average user age (when the data was crawled) was between 40s and 50s, which gives an indication that older people are more interested in book reading when compared to young ones. Also, it looks that people from USA do more social activities than people from other countries. The same point was pointed out by [19]. In addition to the out-degree measure, we evaluated the in-degree measure. The highest and lowest in-degrees and in-degree centralization values of the mother-network were as follows:

TABLE V. HIGHEST AND LOWEST IN-DEGREES AND IN-DEGREE CENTRALIZATION OF THE MOTHER NETWORK

| Metric | Value | Frequency |
|---|---|---|
| Highest input degree value | 707 | 1 |
| Lowest input degree value | 1 | 129480 |
| Network in-degree Centralization | 0.00267556 | - |

We can see that only one node has acquired the highest number of incoming arcs (in-degree) from among 263631 nodes, and that 129480 other nodes acquired only 1 incoming arc.

We can see that nodes (which gained only 1 vote from users for each) represent about half the mother-network. The analysis also gave us 77798 nodes with zero incoming ties (not shown in the table above). This is because the analysis comprised both types of nodes, namely users and books, and nodes with in-degree=0 represent users (source of relation). We can determine the ten books that obtained the highest number of ratings (over the entire rating scale) as follows:

TABLE VI. HIGHEST TEN IN-DEGREE VALUES (REPRESENTING THE BOOKS THAT OBTAINED THE HIGHEST NUMBER OF RATINGS) IN THE BX-DATASET

| Rank | In-degree | Normalized in-degree | ISBN | Book Title |
|---|---|---|---|---|
| 1. | 707 | 0.0027 | 0316666343 | The Lovely Bones |
| 2. | 581 | 0.0022 | 0971880107 | Wild Animus |
| 3. | 487 | 0.0018 | 0385504209 | The Da Vinci Code |
| 4. | 383 | 0.0015 | 0312195516 | The Red Tent (Bestselling Backlist) |
| 5. | 333 | 0.0013 | 0679781587 | Memoirs of a Geisha* |
| 6. | 320 | 0.0012 | 0060928336 | Divine Secrets of the Ya-Ya Sisterhood |
| 7. | 315 | 0.0012 | 059035342x | Harry Potter and the Sorcerer's Stone (Harry Potter (Paperback)) |
| 8. | 307 | 0.0012 | 0142001740 | The Secret Life of Bees |
| 9. | 295 | 0.0011 | 0446672211 | Where the Heart Is (Oprah's Book Club (Paperback)) |
| 10. | 282 | 0.0011 | 044023722x | A Painted House |

The novel 'The lovely bones' has occupied position #1. This is due to the fact that it gained the highest number of users' evaluation and attention. Other books information was taken from the dataset. However, for the ISBN in position 5, we did not find the corresponding information so; we took help from Amazon.com to get the book title and other information. This is an example of the bugs existing in this dataset.

*4.4.2 User-Preference Network Analysis*

This network comprises ratings of users who have rated items with values from 6 to 10 on the scale. The basic idea behind the formation of this network is that our interest is to know whether a user recommends reading/buying a book or not, which means constructing a network of 'likes' and 'dislikes' [17], [34]. However; [19] considered only ratings with 7 or more on the rating scale as positive. Analyzing the network helped us answering the question: which books were most positively-rated (most popular books)?

Let's have a look at some overall statistics of the user-preference network:

TABLE VII. OVERALL STATISTICS OF THE USER-PREFERENCE NETWORK

| Metric | Value |
|---|---|
| Graph Type | Directed |
| Dimension | 228970 |
| Number of Arcs | 363258 |
| Network Density | 0.00000693 |
| Number of Loops | 0 |
| Number of Multiple Lines | 0 |
| Average Degree | 3.17297463 |
| Connected Components | 13979 |
| Single-Vertex Connected Component | 0 |
| Maximum Vertices in a Connected Component | 196180 (85.679%) |

It is a two-mode network consisting of 228970 nodes and 363258 arcs with no edges, since it is a relationship between a user and the book that he/she evaluates. Even though the network density is low (0.00000693), it is still higher than the mother network. This is because the current network has a less number of nodes, as the largest the number of nodes is, the lowest the density. The largest component in this network occupies about 85.679% of the total size of the network. The highest and lowest in-degree values and the network in-degree centralization were as follows:

TABLE VIII.  HIGHEST AND LOWEST IN-DEGREES AND IN-DEGREE CENTRALIZATION OF THE USER-PREFERENCE NETWORK

| Metric | Value | Frequency |
| --- | --- | --- |
| Highest input degree value | 663 | 1 |
| Lowest input degree value | 1 | 112010 |
| Network in-degree centralization | 0.00288867 | |

We can see that nearly half of the user-preference network nodes (i.e. 112010 nodes) obtained only 1 vote, and that only one node obtained the highest number of votes, namely 663.

We can also calculate the highest ten in-degree values (representing most popular books) as follows:

TABLE I.  TOP TEN MOST POPULAR BOOKS IN THE BX-DATASET

| Rank | In-degree | Normalized in-degree | ISBN | Book Title |
| --- | --- | --- | --- | --- |
| 1. | 663 | 0.0029 | 0316666343 | The Lovely Bones |
| 2. | 452 | 0.0020 | 0385504209 | The Da Vinci Code |
| 3. | 344 | 0.0015 | 0312195516 | The Red Tent (Bestselling Backlist) |
| 4. | 307 | 0.0013 | 0679781587 | Memoirs of a Geisha |
| 5. | 305 | 0.0013 | 059035342x | Harry Potter and the Sorcerer's Stone (Harry Potter (Paperback)) |
| 6. | 292 | 0.0013 | 0142001740 | The Secret Life of Bees |
| 7. | 285 | 0.0012 | 0060928336 | Divine Secrets of the Ya-Ya Sisterhood |
| 8. | 274 | 0.0012 | 0446672211 | Where the Heart Is (Oprah's Book Club (Paperback)) |
| 9. | 260 | 0.0011 | 0452282152 | Girl with a Pearl Earring |
| 10. | 250 | 0.0011 | 0671027360 | Angels & Demons |

The table above lists the ten most popular books. The more in-degree value is, the more prestigious the book. With this metric, we can say that the most preferred (popular) book (at the time when the data was crawled) by users was "The Lovely Bones: A novel".

*4.4.3  User Non-Preference Network Analysis*

The third network that we analyzed was the user non-preference network. It comprised users who have rated books with values from 1 to 5 on the rating scale. Analyzing the network helped us answering the question: which books were most negatively-rated (most un-popular books)?

TABLE IX.  OVERALL STATISTICS OF THE USER NON-PREFERENCE NETWORK

| Metric | Value |
| --- | --- |
| Graph Type | Directed |
| Dimension | 73716 |
| Number of Arcs | 70403 |
| Network Density | 0.00001296 |
| Number of Loops | 0 |
| Number of Multiple Lines | 0 |
| Average Degree | 1.91011449 |
| Connected Components | 10865 |
| Single-Vertex Connected Component | 0 |
| Maximum Vertices in a Connected Component | 45008 (61.056%) |

It is a two-mode network consisting of 73716 nodes and 70703 arcs with no edges or loops. Network Density = 0.00001296 which is very low (however, it is still higher than the two previous networks since this network has only 73716 nodes). We notice that the number of nodes here exceeds the number of arcs, which indicates users' less interest to evaluate books if they did not like. The number of connected components and the average degree are less than its two previous networks (Tables II, VII). It has less average degree value because the number of arcs here is less than the number of nodes.

The highest and lowest in-degree values and the network in-degree centralization were as follows:

TABLE X.  HIGHEST AND LOWEST IN-DEGREES AND IN-DEGREE CENTRALIZATION OF THE USER NON-PREFERENCE NETWORK

| Metric | Value | Frequency |
| --- | --- | --- |
| Highest input degree value | 389 | 1 |
| Lowest input degree value | 1 | 41447 |
| Network Input Degree Centralization | 0.00526420 | - |

More than half of the network nodes (books) obtained only 1 vote for each, while the highest in-degree value in the user-non preference network was 389, which means that the corresponding book was rated by the users as the most unpopular book.

By implementing the in-degree measure, we get the following ten results which represent most unpopular books:

TABLE XI.  TOP IN-DEGREE VALUES (REPRESENTING MOST UNPOPULAR BOOKS) OF THE USER NON-PREFERENCE NETWORK

| Rank | In-degree | Normalized in-degree | ISBN | Book Title |
| --- | --- | --- | --- | --- |
| 1. | 389 | 0.0053 | 0971880107 | Wild Animus |
| 2. | 51 | 0.0007 | 044023722x | A Painted House |
| 3. | 44 | 0.0006 | 0316666343 | The Lovely Bones |
| 4. | 41 | 0.0006 | 0316601950 | The Pilot's Wife |
| 5. | 41 | 0.0006 | 0316769487 | The Catcher in the Rye |
| 6. | 39 | 0.0005 | 0312195516 | The Red Tent (Bestselling Backlist) |
| 7. | 39 | 0.0005 | 0446605239 | The Notebook |
| 8. | 38 | 0.0005 | 0425182908 | Isle of Dogs |
| 9. | 36 | 0.0005 | 0140293248 | The Girls' Guide to Hunting and Fishing |
| 10. | 35 | 0.0005 | 0375727345 | House of Sand and Fog |

We notice that two of the books in the table above (positions 3 and 6) have also been seen in the user-preference network (Table I). This may reflect the fact that users' choices covered a wide range of ratings over a scale (from 1 to 10), and that peoples' opinions towards these books largely scattered between "good" and "bad".

*4.5 Affiliation Network Analysis*

The term Affiliation refers to membership or participation data such as when we have data on which actors have participated in which events. It can be represented as a bipartite graph (*V1, V2, E*), where *V1* and *V2* are two different sets of nodes, while *E* is an affiliation relation between elements of *V1* and *V2* [2].

Usually, we can extract two one-mode networks from one a two-mode network as follows: the first one is the network of interlocking events (if two books share the same event i.e. being read by the same two or more readers) and the second one is the network of actors (if two users or more like the same books). The idea behind inducing co-affiliation network from affiliation network is that a co-affiliation network provides the ground for the development of social relationships between the actors of one set. For example, the more the number of times people come at the same event, the more likely those people are going to interact and develop some type of relationship. It has been reported that persons whose activities are focused around the same point, frequently become connected over time.

*4.5.1 User-User Network Analysis*

For the purpose of affiliation network analysis, we made use of the lately generated user-preference network to generate this new network which will help us later on probing the potential social relations among users. It is a network with connections between users only.

We restricted ourselves here to extract this network from the user-preference network (rather than other networks), because what makes people develop friendships depends mainly on the things they share and the things they like.

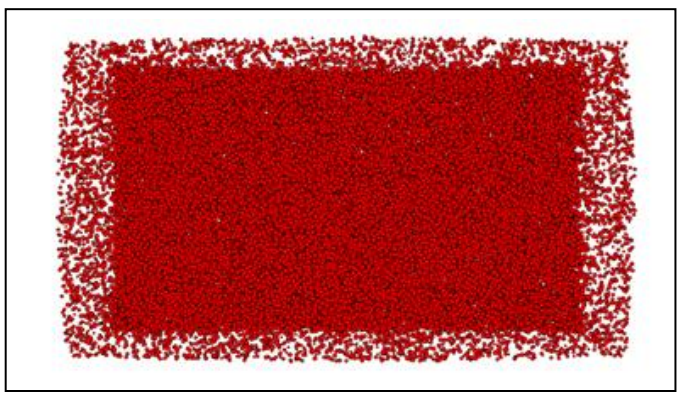

Figure 1.  2-D representation of the user-user network.

Figure (1) gives a 2-D representation of the user-user network. The network was energized using Fruchterman-Reingold algorithm [9]. Edges and vertex labels have been eliminated. Nodes in the middle are the core nodes, while nodes around the core are the periphery nodes.

Some overall statistics of the user-user network are as follows:

TABLE XII. OVERALL STATISTICS OF THE USER-USER NETWORK

| Metric | Value |
|---|---|
| Graph Type | Undirected |
| Dimension | 69768 |
| Number of Edges | 3176585 |
| Network Density | 0.00130522 |
| Number of Loops | 0 |
| Number of Multiple Lines | 0 |
| Connected Components | 883 |
| Single-Vertex Connected Component | 13096 |
| Maximum Vertices in a Connected Component | 54701(78.404%) |
| Maximum Geodesic Distance (Diameter) | 10 |
| Average Geodesic Distance (Among Reachable Pairs) | 2.80782 |
| Average Degree | 91.06137484 |
| Number of Unreachable Pairs | 1875356164 |

It is a one-mode undirected sub-network consisting of 69768 vertices and 3176585 weighted edges. Network density =0.001305 which is higher than the earlier networks. This is because a 1-mode network has higher density than its equivalent a 2-mode network since in 1-mode network; vertices can have ties with any other nodes in the network, while this is not true for 2-mode networks.

The results showed that for (n>=2), the network consisted of 883 components. The size of the largest component is 54701 (78.404%), while the size of the next largest component is 13096 (18.770%, not shown here) and the rest of components constitute approximately 10% of the network. Network diameter, which is the longest shortest path in the network, is 10. This geodesic distance exists only between two users, namely 150578 and 112131. The first guy is 43 years old from Milano, Italy while the other guy is 12 years old from Sydney, Australia. This could be due to variation in age and the geographic locations of both.

We can see that at this time, the network not only having connected components of two or more vertices, but also having single-vertex connected components = 13096, which means that it includes 'isolates'. This is because the user-user sub-network emerges from a larger network, namely the user-preference network, which already contains books having in-degree value=1. When extracting a one-mode subnetwork from a two-mode network, these nodes become 'isolates'. The network has 1875356164 unreachable pairs, which expresses the number of pairs of nodes that do not have a connection between them.

*4.5.2 Applying Centrality Measures*

We want to infer the most potential central people in the user–user network. So, we are going to implement the three measures of centrality, namely degree, closeness and betweenness centrality measures. Research has proved that these three measures are highly correlated and give similar results in identifying most important actors in a network [3]. The importance behind identifying most important actors is that it reflects how active an actor is. Also, active actors are more likely to establish social ties with a large number of other actors and can affect how the network works.

First, we are going to find top-degree centrality users using the in-degree measure. Figure (1) was built based on node (circle) size. The larger the node is, the more central a user in the network in regard to degree centrality.

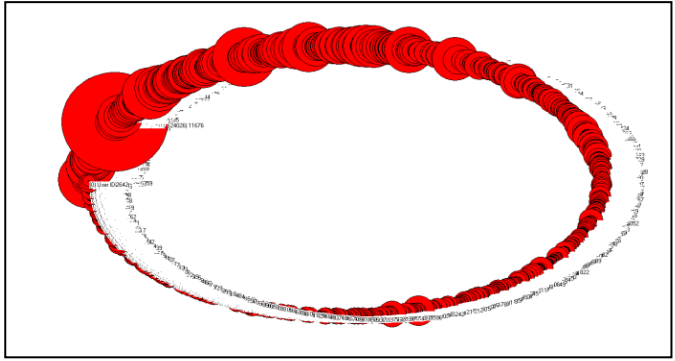

Figure 2. Circular 2-D representation of the degree-centrality measure in the user-user network

Degree centrality statistics of the user-user network were as follows:

TABLE XIII. DEGREE CENTRALITY STATISTICS OF THE USER-USER NETWORK

| Metric | Value |
|---|---|
| Dimension | 69768 |
| Highest degree centrality value | 24026 |
| Lowest degree centrality value | 0 |
| Network Input Degree Centralization | 0.34307946 |

We can see that we have nodes with degree centrality =0 because these are 'isolates'. Highest ten degree centrality values in the user-user network were as follows:

TABLE XIV. HIGHEST TEN DEGREE CENTRALITY VALUES IN THE USER-USER NETWORK

| Rank | User ID | Degree Centrality | Demographic Info |
|---|---|---|---|
| 1. | 11676 | 24026 | N/A |
| 2. | 16795 | 8614 | Mechanicsville, Maryland, USA, 47 Years Old |
| 3. | 95359 | 8110 | Charleston, west Virginia, USA, 33 Years Old |
| 4. | 60244 | 6493 | Alvin, Texas, USA, 47 Years Old |
| 5. | 204864 | 6104 | Simi valley, California, USA, 47 years Old |
| 6. | 104636 | 5533 | Youngstown, Ohio, USA |
| 7. | 98391 | 5480 | Morrow, Georgia, USA, 52 years old |
| 8. | 35859 | 5409 | Duluth, Minnesota, USA |
| 9. | 135149 | 5283 | ft. Pierce, Florida, USA |
| 10. | 153662 | 5281 | ft. Stewart, Georgia, USA, 44 years old |

The user with the highest degree centrality was #11676. However, we didn't find any demographic information related to him/her, as it seems he/she preferred to keep identification information dim. That guy has already occupied position #1 in terms of people with the highest number of outgoing ties in the mother-network (Table IV).

That guy has the largest potential social network, as he/she is connected in a direct path to 24026 other actors (neighbors) in the network, which means that he/she shares common opinions about a specific number of book(s) with other 24026 users in the user-user network. This high number of connections reflects the fact that a 1-mode network has a higher density than a 2-mode network as nodes can freely connect to any other nodes in the same network.

The user in position #2, namely user ID 16795 (47 years old of Maryland, USA), has the second largest potential social network consisting of 8614. However, he/she came only in rank #9 in a previous statistics about users with the highest number of outgoing ties (Table IV). This might mean that even though that guy had fewer number of outgoing ties than the other eight guys, his/her choices were more focused and that he/she could share book preferences with more other people, which makes him/her more candidate to establish social relations than others (of course behind our top-user).

The second measure of centrality, we use here, is closeness. The concept of closeness centrality depends on the total distance between one vertex and all other vertices, as large distances show lower closeness centrality. Closeness centrality values range from 0 (for isolated vertices) to 1. For a specific vertex, it results from the number of all other vertices in the network divided by the sum of distances between that vertex and all other vertices in the network. Therefore, closeness centrality values are continuous rather than discrete [5].

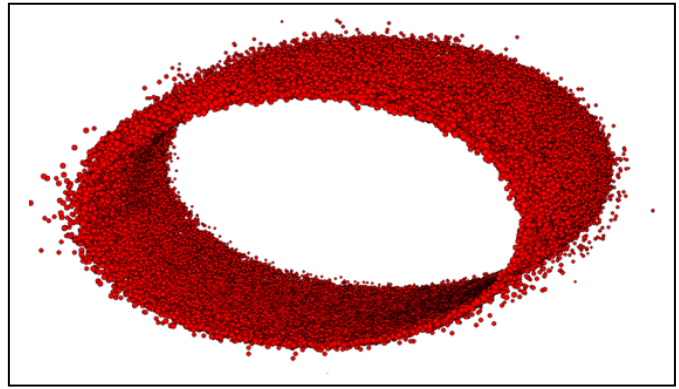

Figure 3. Circular 3-D representation for the closeness centrality measure of the user-user network. All nodes are equally sized

It took about 15 hours to calculate closeness centrality values of all vertices in the network however; it depends mainly on the device specifications. Some overall statistics for closeness centrality are as follows:

TABLE XV. OVERALL STATISTICS FOR CLOSENESS CENTRALITY MEASURE IN THE USER-USER NETWORK

| Metric | Value |
|---|---|
| Dimension | 69768 |
| Highest closeness centrality value | 0.4926 |
| Lowest closeness centrality value | 0.0000 |
| Arithmetic mean | 0.2233 |
| Median | 0.2661 |
| Standard deviation | 0.1220 |
| Network closeness centralization cannot be computed since the network is weakly connected | - |

The closeness centrality values of the first ten actors were as follows:

TABLE XVI. CLOSENESS CENTRALITY VALUES OF THE FIRST TEN ACTORS IN THE USER-USER NETWORK

| Rank | Closeness Centrality | User ID | Demographic Info |
|---|---|---|---|
| 1. | 0.4926 | 11676 | N/A |
| 2. | 0.4021 | 16795 | Mechanicsville, Maryland, USA, 47 years old |
| 3. | 0.4011 | 95359 | Charleston, west Virginia, USA, 33 years old |
| 4. | 0.3919 | 60244 | Alvin, Texas, USA, 47 years old |
| 5. | 0.3906 | 204864 | Simi valley, California, USA, 47 years old |
| 6. | 0.3862 | 35859 | Duluth, Minnesota, USA |
| 7. | 0.3860 | 135149 | ft. Pierce, Florida, USA |
| 8. | 0.3852 | 104636 | Youngstown, Ohio, USA |
| 9. | 0.3850 | 153662 | ft. Stewart, Georgia, USA, 44 years old |
| 10. | 0.3838 | 98391 | Morrow, Georgia, USA, 52 years old |

It is easy to notice that the user (id=11676) is the top-closeness centrality user. This is mainly true because he/she is the top out-degree user (Table: IV), and the top in-degree user (Table: XIV). The rest of actors in the table also appeared in the study, which reflects their importance at the social level, alongside the ultimate importance of the top-user (namely user ID= 11676). Network closeness centralization cannot be computed if the network was not strongly connected since there are no paths between all vertices so; it is impossible to compute the distances between some vertices [5].

While degree and closeness centrality are based on the concept of the reachability of a person, betweenness centrality is based on the idea that a person is more important if he/she was more intermediary in the network. The more a person is a go-between, the more central her/his position in that network. This reflects the importance of a person being in the middle of social communications of a network and to what extent he/she is needed as a link in the chains of contact in the society. On the other hand, a vertex has betweenness centrality = 0 if it was not located between any other vertices in the network, which points out to a weak social role that he/she plays. Many vertices may not appear in the figure below (Figure: 4) because they do not mediate between any two vertices, so their betweenness centralities equal zero. The drawing was built based on node size. The larger the node (circle) is, the more central the user in the network, in regard to betweenness centrality concept.

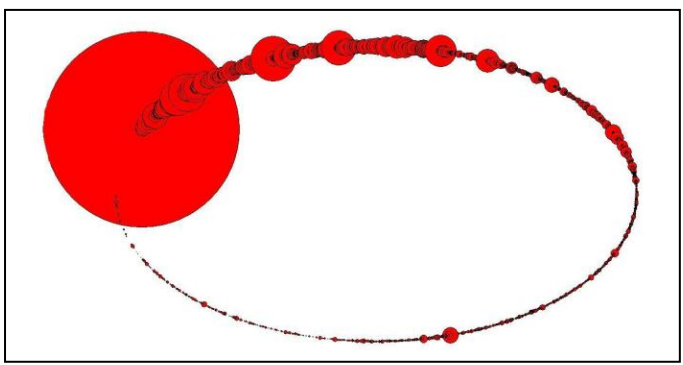

Figure 4. Circular representation for betweenness-centrality measure of the user-user network

It took about 15 hours to calculate all betweenness centrality values. Some overall statistics of the betweenness centrality measure were as follows:

TABLE XVII. SOME OVERALL STATISTICS OF THE BETWEENNESS CENTRALITY MEASURE IN THE USER-USER NETWORK

| Metric | Value |
|---|---|
| Dimension | 69768 |
| Highest betweenness centrality value | 0.1735 |
| Lowest betweenness centrality value | 0.0000 |
| Arithmetic mean | 0.0000 |
| Median | 0.0000 |
| Standard deviation | 0.0007 |
| Network betweenness centralization | 0.17345915 |

The betweenness centralities of the first ten actors in the network were as follows:

TABLE XVIII. TOP BETWEENNESS CENTRALITY MEASURE VALUES IN THE USER-USER NETWORK

| Rank | Betweenness Centrality | User ID | Demographic Info |
|---|---|---|---|
| 1. | 0.1735 | 11676 | N/A |
| 2. | 0.0121 | 98391 | Morrow, Georgia, USA, 52 Years Old |
| 3. | 0.0094 | 16795 | Mechanicsville, Maryland, USA, 47 Years Old |
| 4. | 0.0085 | 95359 | Charleston, west Virginia, USA, 33 Years Old |
| 5. | 0.0065 | 153662 | ft. Stewart, Georgia, USA, 44 Years Old |
| 6. | 0.0055 | 204864 | Simi valley, California, USA, 47 years old |
| 7. | 0.0055 | 60244 | Alvin, Texas, USA, 47 years old |
| 8. | 0.0053 | 23902 | London, England, United Kingdom |
| 9. | 0.0047 | 135149 | ft. Pierce, Florida, USA |
| 10. | 0.0045 | 104636 | Youngstown, Ohio, USA |

The top-user (user id=11676) is still in rank #1 in the table which means that he/she lies at the geodesic distances between other pairs, more than any other vertex in the network. This nominates him/her (more than others) to be a candidate person to play many potential brokerage roles in the future.

As we notice, all the three measures (degree, closeness and betweenness centrality) have showed similar (not identical) results, which support the notion that all these measures collectively are used to measure most important individuals in a community

### 4.5.3 Ego-Network Analysis

After conducting a comprehensive analysis using some important measures in SNA, we turn our eyes to the top-user (ID = 11676) who occupied the first position in all the previous tests (Tables: IV, XIV, XVI, XVIII), and try to analyze his/her sub-network (which is called ego-network or ego-centric approach as opposed to the socio-centric approach).

A very useful way to understand complicated networks is to see how they arise from the local connections of individual actors.

Ego-network (which consists of ego, its neighbors and ties among them) was extracted the from the user-user network. We show below a 2-D representation of the ego-network:

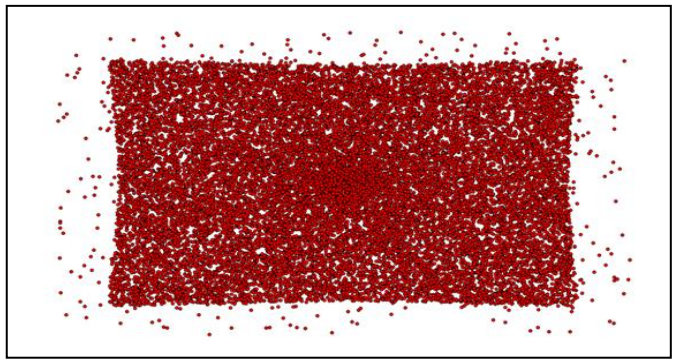

Figure 5. 2-D Representation of the ego-network, extracted from the original user-user network

Let's take a look at a short summary of some metrics, evaluated using Pajek:

TABLE XIX. SUMMARY OF THE EGO-NETWORK STATISTICS

| Metric | Value |
|---|---|
| Graph Type | Undirected |
| No. of Neighbors | 24026 |
| Number of Edges | 2278058 |
| Ego-network Density | 0.00789314 |
| Number of Loops | 0 |
| Number of Multiple Lines | 0 |
| Maximum Geodesic Distance (Diameter) | 8 |
| Average Geodesic Distance | 2.49241 |
| Average Degree | 189.63273121 |
| Ego-network Betweenness Centralization | 0.02163385 |

The network consists of 24026 neighbors. Those neighbors are only the direct ones, i.e. who are located at distance one from ego. Also, the number of edges is 2278058. This number represents the relations among vertices around ego.

The density of ego-network expresses the density of ties among its neighbors. The result is 0.00789314 which is relatively high and at the same time higher than the densities of our earlier networks (namely the mother, the user-preference, the user non-preference and the user-user networks), which means that ego-network is quite embedded in dense local substructure. This is because this network is the local network of the top-user (who occupied the 1st position in all the previous four tests). The ego network diameter and the average geodesic distance are a little slighter than the user-user network (Table XII). This is intuitive since the current network is a dense fragment of the user-user network.

Ego-network diameter, that is the maximum geodesic distance between two vertices, is 8. This geodesic exists between User ID=47534 (45 years from Luzern, Switzerland) and User ID=240418 (34 years old from Barcelona, Spain).

Also, we notice that ego-network betweenness centralization is 0.02163385 which is lower than what it is in the user-user network (Table XVII).

This is because the variation in vertex betweenness centrality in the user-user network is higher than what it is in ego-network (Figure: 6).

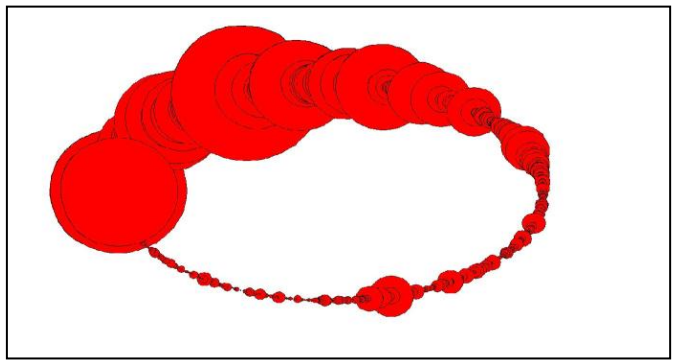

Figure 6. Ego-network betweenness centralization

We can calculate the geodesics from the top-user to all other vertices in the user-user network as follows:

TABLE XX. GEODESIC DISTANCES FROM TOP-USER TO ALL OTHER USERS IN THE USER-USER NETWORK

| Cluster | Frequency |
|---|---|
| 0 | 1 |
| 1 | 24026 |
| 2 | 29088 |
| 3 | 1490 |
| 4 | 86 |
| 5 | 10 |
| Sum | 54701 |
| Unknown | 15067 |
| Total | 69768 |

We can see that the top-user can reach 24026 users with only one hop. He/she can reach 29088 other users with two hops, 1490 others with three hops and so on.

However, there are other 15067 vertices that can't be reached by ego; in other word they are unreachable in our ego-network, and hence are given the value 999999997 in Pajek's report of distances. In the ego-network, there are unreachable nodes because our user-user network is split up into smaller parts (883 connected components. Table: XII). Cluster (0) means that ego doesn't need any hop to get to that node, as it is the ego him/her-self.

We can also calculate the potential brokerage roles practiced by ego in the user-user network. Brokerage expresses the ability to induce and exploit competition between the other two actors of the triad (a triad consists of a focal person, alter and a third person in addition to the ties among them), and also expresses his/her qualifications to play a subversive role through creating or exploiting conflict between the other two actors in order to control them [5].

Brokerage can be calculated by using the 'aggregate constraint' concept which is the sum of the dyadic constraint on all of a vertex's ties. However, the aggregate constraint has an opposite effect, i.e. the more the aggregate constraint, the less the brokerage role an actor can play.

The implementation gave us the distribution table of aggregate constraints. We put down here the two extremes:

TABLE XXI. THE TWO EXTREMES OF AGGREGATE CONSTRAINT IN THE USER-USER NETWORK

| Aggregate Constraint | Value | Representative |
|---|---|---|
| Highest Value | 1.3203 | 44726 |
| Lowest Value | 0.0007 | 11676 |

We see that the top-user (ego) has the lowest aggregate constraint in the user-user network because he/she has the highest out-degree in the mother network (Table: IV), in-degree, betweenness and closeness values in the user-user network (Tables: XIV, XVI, XVIII). In other words; he/she can perfectly play brokerage.

### 4.5.4 M-Slice Analysis

A one-mode network induced from a two-mode network creates the atmosphere to discover many dense structures. One way to detect cohesive subgroups in one-mode networks is to detect m-slice sub-networks. M-slice can be defined as the maximal sub-network in which line multiplicity is equal or greater than m. It was first introduced by John Scott as 'm-core'. This technique puts into consideration line multiplicity rather than the number of neighbors (which is defined by the k-core concept). M-slice method comprises allocating values to network nodes based on m-slice, i.e. the highest tie these nodes are incident (connected) with. The importance of conducting this type of analysis is that it helps us identify the strongest potential social relations in the network based on 'participation rate' between each pair of nodes. It has been found that the larger the number of interlocks between two users, the stronger their tie (or relationship) and the more similar they are [5]. We first examine the network in order to find out the distribution of tie weights, as these weights control how m-values are allocated to nodes:

TABLE XXII. DISTRIBUTION OF TIE WEIGHTS IN THE USER-USER NETWORK

| I | Tie Weights | Frequency |
|---|---|---|
| 1 | 36.0000 | 24834 |
| 2 | 36.0000 - 8470.3333 | 3151746 |
| 3 | 8470.3333 - 16904.6667 | 4 |
| 4 | 16904.6667 - 25339.0000 | 1 |
|   | Total No. of Links | 3176585 |

The results above show that the lowest line multiplicity is 36 (achieved in 24834 ties) and the highest line multiplicity is 25339 (achieved in only 1 tie). From the m-slice frequency tabulation values of the user-user network, we display the highest five values in addition to the lowest five values:

TABLE XXIII. M-SLICE VALUES IN THE USER-USER NETWORK

| M-slice | Value | Number of Nodes | Representative |
|---|---|---|---|
|  | 0 | 13096 | - |
|  | 36 | 160 | - |
| Lowest five values | 42 | 525 | - |
|  | 48 | 774 | - |
|  | 49 | 621 | - |
|  | 25339 | 2 | 98391, 235105 |
|  | 16129 | 1 | 11676 |
| Highest five values | 9864 | 1 | 153662 |
|  | 9466 | 1 | 16795 |
|  | 7781 | 1 | 104636 |

The results above show that 13096 of the nodes belong to the 0-slice, which means that these nodes are not connected to any nodes in the network and that the users do not share book preference among them or with any other users. In other words; they are 'isolates'. They represent the weakest potential social components in the user-user network. In fact, they constitute no social components at all (in the context of our measures). It is not likely that those users in the future establish relationships among them by any means or of any type, since there is nothing they can gather on.

We can see also that the strongest potential social component (which belongs to the 25339-slice) consists of two nodes: 98391 (52 years old from Georgia, USA) and 235105 (46 years old from Missouri, USA). This pair can formulate the most powerful, everlasting, and fast-shaping relationship. Many reasons may stand behind that, for instance: age, occupation, level of education, environment, gender, past experience, marital status and so on.

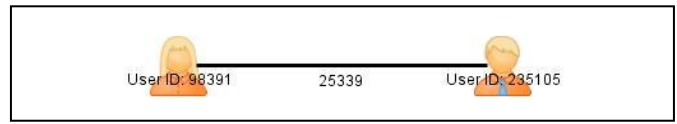

Figure 7. The strongest potential social component in the user-user network which comprises two vertices (98391 and 235105)

Using the m-slice concept, we can extract stronger and stronger subgroups by removing undesired lines and nodes that do not satisfy our goals. The process will raise the minimum m-slice threshold, which in turn forms more cohesive groups.

For example, if we remove the 0-slice nodes, the resulting network will consist of 56672 and 3176585 lines. Next, we need to eliminate unnecessary lines. Thus, we obtain more cohesive components. If we keep going on that process, we will end up with the highest m-slice component, namely 25339-slice that consists of only two nodes (98391, 235105).

### 5. CONCLUSION AND FUTURE WORK

The purpose of this study was to present an in-depth analysis of one of the most celebrated social sites frequently visited by individuals who are interested in exchanging information about the books they have already experienced. The research questions of interest were addressed via analyzing the relational data to detect social interaction schemes and to find out most celebrated features that characterize this community.

In order to narrow the results above, we calculated the number of appearance of each user (in each category). For example, the top-user (user id=11676) appeared in four places so, he/she is in category #4 (A category represents the number of appearance for each user within the list of ten top-users). Also, he/she has occupied the first positions in all these four tests so; he/she gets four points (by multiplying 1*4).

This user is in a better position as compared to his subsequent fellow (in the below table), namely user id=16795, who belongs to category #4 also but obtained 16 points (9+2+3+2=16). As a rule of thumb, we shall suppose: the less the number of points is, the higher the rank of a user. The overall top ten users within the Book-Crossing (at the time when that data was crawled) were as follows:

TABLE XXIV. TOP TEN USERS WITHIN THE BOOK-CROSSING DATASET

| Rank | User ID | Category | Points | Demographic information |
|---|---|---|---|---|
| 1. | 11676 | 4 | 4 | Null |
| 2. | 16795 | 4 | 16 | Mechanicsville, Maryland, USA, 47 Years Old |
| 3. | 98391 | 4 | 21 | Morrow, Georgia, USA, 52 years old |
| 4. | 153662 | 4 | 27 | ft. Stewart, Georgia, USA, 44 years old |
| 5. | 95359 | 3 | 10 | Charleston, west Virginia, USA, 33 years old |
| 6. | 60244 | 3 | 15 | Alvin, Texas, USA, 47 years old |
| 7. | 204864 | 3 | 16 | Simi valley, California, USA, 47 years old |
| 8. | 104636 | 3 | 24 | Youngstown, Ohio, USA |
| 9. | 135149 | 3 | 25 | ft. Pierce, Florida, USA |
| 10. | 23902 | 2 | 13 | London, England, United Kingdom |

The results show that 8 to 9 of actors were from USA and that only 1 to 2 of actors was from a country rather than USA, namely United Kingdom. The results also show that almost half of the actors were in 40s. The lack of more demographic information has stopped us from knowing more about the implications behind users' choices.

For the books that earned the highest number of ratings, whether they were negative or positive (from 1 to 10 on the rating scale), we obtained the results showed in Table: (VI). Although these books obtained a large number of users' evaluations, they are not necessarily considered the most preferable books to users. We can say these books took a wide range of users' interest, and that users had different impressions about these books which in turn pushed them to take different perspectives.

The books that earned the highest number of positive ratings (from 6 to 10 on the rating scale) were showed in Table: (I). Eight of these books also appeared within the list of the books that earned the highest number of ratings (Table: VI) and that some of them have become the story of cinema movies (e.g. the Da Vinci Code). Also, the author "Dan Brown" had two books within this list, namely the book in position #2 and in position #10. This may reflect his significance as a key author in the world of books.

For the books that earned the highest number of negative ratings (i.e. the unpopular books), we obtained the results showed in Table: (XI). We can see that 2 of these books also appeared in the list of books that earned the highest number of positive ratings (namely books in position #3 and position #6). This gives an indication that users' opinions towards these books scattered across the entire scale and that people were inconsistent about them.

The research also took us to dig out the most powerful and the weakest social relationships within the hypothetical user-user network by using m-slice type of analysis (Table: XXIII).

We can see that the weakest relationships have weight=0, which means that these entities represent isolated nodes. The number of weakest relationships is 13069 relationships (nodes). Also, the strongest potential relationship has a weight=25339 and that only one entity represents this relationship, which exists between two nodes (98391 & 235105).

The research methodology of this study can be further extended to other online social networks rather than the book-Crossing community. Any website where people are able to rate items on a specific scale (e.g. from 1 to 5 or 10) will be a good place to induce potential social relations from that community. Many websites, these days, give the space for their visitors to rate the materials they have bought or only checked.

We can build map of user preferences which will help us further predict user behaviors and even give recommendations to similar associates (based on either most important people in the network or through the help of m-slice analysis).

We can further make the process more autonomous and develop an agent that can automatically visit a specific website and recursively extract huge amount of data (maybe bigger than the current one). But we should keep in mind at the same time preserving user privacy.